\documentclass[twocolumn,showpacs,preprintnumbers,amssymb]{revtex4}
\usepackage{pstricks}%
\usepackage{graphicx}
\usepackage{dcolumn}
\usepackage{bm}
\usepackage{dsfont}
\usepackage{amsmath}
\usepackage{pifont}
\usepackage{psfrag}
\usepackage{hyperref}
\usepackage[latin2]{inputenc}
\usepackage{bbm}

\renewcommand\Re{\text{Re\,}}
\renewcommand\Im{\text{Im\,}}
\newcommand\Tr{\text{Tr\,}}

\newcommand{\BE}{\begin{equation}}
\newcommand{\EE}{\end{equation}}

\newcommand{\skipc}[2]{}

\newcommand{\tab}[1]{Tab.~\ref{#1}}
\newcommand{\fig}[1]{Fig.~\ref{#1}}
\newcommand{\eq}[1]{Eq.~(\ref{#1})}
\newcommand{\Sec}[1]{Sec.~\ref{#1}}
\newcommand{\App}[1]{App.~\ref{#1}}

\newcommand{\I}{\ensuremath{{\mkern1mu\mathrm{i}\mkern1mu}}}
\newcommand{\E}{\ensuremath{{\mkern1mu\mathrm{e}\mkern1mu}}}

\newcommand{\qed}{\nobreak \ifvmode \relax \else
      \ifdim\lastskip<1.5em \hskip-\lastskip
      \hskip1.5em plus0em minus0.5em \fi \nobreak
     $\square$\fi}

\newcommand{\be}{\begin{equation}}
\newcommand{\ee}{\end{equation}}
\newcommand{\eea}{\end{eqnarray}}
\newcommand{\bea}{\begin{eqnarray}}

\newcommand{\ket}[1]{\ensuremath{|#1\rangle}}
\newcommand{\bra}[1]{\ensuremath{\langle#1|}}

\newcommand{\map}{\mathcal{E}}

\usepackage{color}

\begin{document}

\title{Distinguishing between statistical and systematic errors in  quantum process tomography}

\author{Sabine W\"olk}
\affiliation{Naturwissenschaftlich-Technische Fakult\"at, Department Physik, Universit\"at Siegen, 57068 Siegen, Germany}
\affiliation{Institute for Theoretical Physics, University of Innsbruck, Technikerstra{\ss}e 21a, 6020 Innsbruck, Austria}
\author{Theeraphot Sriarunothai, Gouri S. Giri and Christof Wunderlich}
\affiliation{Naturwissenschaftlich-Technische Fakult\"at, Department Physik, Universit\"at Siegen, 57068 Siegen, Germany}

\date{\today}

\begin{abstract}
It is generally assumed that every process in quantum physics can be described mathematically by a completely positive map. However, experimentally reconstructed processes are not necessarily completely positive due to statistical or systematic errors. In this paper, we introduce a test for discriminating statistical from systematic errors which is necessary to interpret experimentally reconstructed, non-completely positive maps. We demonstrate the significance of the test using several examples given by experiments and simulations. In particular, we demonstrate experimentally how an initial correlation between the system to be measured and its environment  leads to an experimentally reconstructed map with negative eigenvalues. These experiments are carried out using atomic $^{171}$Yb$^+$ ions confined in a linear Paul trap, addressed and coherently manipulated by radio frequency radiation.  
\end{abstract}


\maketitle

\section{Introduction}


The time evolution of a state $\rho_S$ of a quantum system is generally described by a completely positive (CP) map $\map$ to ensure that positive quantum states stay positive. Yet, maps reconstructed via experimental process tomography often tend to be not completely positive \cite{Weinstein2004, Ringbauer2015, Kuah2007, Pechukas1994, Wood2009,Modi2010,Ziman2006, Knips2015}. 

There exist several reasons for the appearance of non-positive maps in quantum process tomography: (i) statistical errors due to limited number of measurements \cite{Knips2015}, or systematic errors such as  (ii) misaligned measurements and preparation errors or (iii) initial correlation between the system and the environment \cite{Pechukas1994, Wood2009,Modi2010,Ziman2006}. Such initial correlation can arise  if the preparation of the system also affects the environment. In the first two cases, the resulting map can be non-positive meaning that the reconstructed state $\rho_S'$ may have negative eigenvalues. Or, $\map$ is positive but not completely positive meaning that $\rho_S'$ itself is positive, but the time evolution of a larger (composite) system with $\map$ acting only on one part of it leads to negative eigenvalues of the state of the total system. These types of errors arise also in quantum state tomography. On the other hand, in the case of (iii) initial correlation between the system and the environment, the resulting map will be positive but not completely positive \cite{Pechukas1994, Wood2009,Modi2010,Ziman2006}. This effect does not arise in state tomography and is therefore a new and unique feature of process tomography.

Furthermore, a general mathematical description of the time evolution of a system does not exist, if the environment is correlated with the system \cite{Carteret2008, Vacchini2016}. In such a system, any time evolution of the state $\rho_S$   is possible which maps valid quantum states to valid quantum states. Therefore, knowing the time evolution of $d^2$ states of a $d$-dimensional Hilbert space, as in process tomography, is not enough to predict the map $\map$ \cite{Modi2012}.

 Nevertheless, one goal in experimental quantum information science is to isolate quantum systems in such a way, that they can be approximated by closed quantum systems. Therefore, the assumption  that quantum channels can be represented by completely positive maps is well justified \cite{Carteret2008} but has to be checked for a given experimental process. 
Therefore, if a non-completely positive map appears in quantum process tomography, it is important to decide whether the negativity is the result of statistical or systematic errors \cite{Schwemmer2015}. In the first case, one may ignore the negativity, or record more data to reduce it. However, in the second case we have to find the error and modify our experiment by either improving our control of the system to reduce preparation and measurement errors or to better isolate our system from the environment.  Indications for systematic errors can be found by just analyzing the collected data without changing the experiment  as we will demonstrate in this paper. The method introduced here  is an important tool for quantum process tomography, since it gives meaningful hints about possible systematic errors at a very low cost in terms of experimental resources. 

Quantum process tomography \cite{Nielsen2000} is an important tool to experimentally verify quantum gates \cite{oBrien2004,Mitchell2003} and to investigate complex quantum systems \cite{Gessner2014}. Quantum process tomography is the most detailed characterization of gates, but it is very resource-intensive \cite{Bagan2003}. The effort can be reduced for matrices with low  rank by using methods from compressed sensing \cite{Gross2010,Shabani2011}. Another possibility is a process certification with the help of the Monte Carlo method \cite{Flammia2011,Silva2011,Steffen2012}. Here, the average output fidelity compares the experimentally realized process with the target unitary. For process tomography of a quantum gate, the gate is applied to $N$ states which are eigenstates of a random combination of local Pauli operators. For each state, the fidelity between the ideal output and the experimentally realized output is determined. With the help of these state fidelities, the average output fidelity between the experimentally realized gate and the ideal gate can by estimated with an uncertainty which decreases as $1/ N$ .

Another way to approximate the quantum process fidelity, suggested by H. Hofmann \cite{Hofmann2005}, uses two sets of mutually unbiased bases. Applying an ideal unitary quantum
gate on each basis leads to an orthogonal output basis, which makes the measurement of the fidelity between the ideal output and the experimentally realized output easy. For each of the two bases, the average state fidelity is calculated, which are upper bounds of the process fidelity. A lower bound is given by the sum of both fidelities minus one. This method was used to characterize a CNOT gate realized with a four-photon six-qubit cluster state \cite{Gao2010}. In a similar fashion,  other properties of channels can also be characterized \cite{Orieux2013}. 

In what follows, we first shortly recapitulate in \Sec{sec:ncp_maps} process tomography and discuss the meaning of negative eigenvalues in the case of initial correlation  between the system and its environment. Then, we explain in \Sec{sec:test} a plausibility check testing the probability, that the non-positivity of a reconstructed quantum process is due to  statistical effects. Consecutively, we test the performance of the introduced consistency test in \Sec{sec:examples}. In \Sec{sec:simulation}, we first present our simulations. Then, we introduce and carry out an experiment where we intentionally engineer an initial correlation between the system and the environment, each given by a single trapped ion , and apply our test to the experimentally reconstructed process (\Sec{sec:experiment}). Finally, we finish this article with conclusions in \Sec{sec:conclusion}.

\section{Process tomography and system environment correlations\label{sec:ncp_maps}}

\begin{figure}[t]
\includegraphics[width=0.35\textwidth]{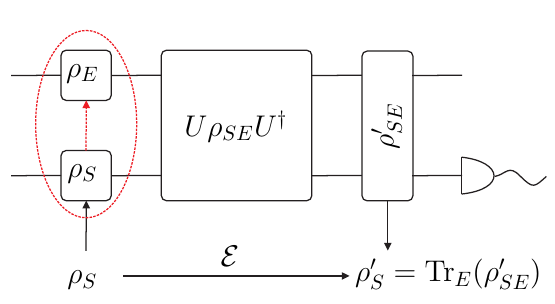}
\caption{Systematic description of the time evolution of an open quantum system. Typically, we assume that the preparation of the system state $\rho_S$ does not influence the initial state of the environment. If there exist initial correlations (represented by red dotted lines), the resulting map $\map$ is not necessarily completely positive. }
\label{fig:time_evo}
\end{figure}

The time evolution of every quantum system can be described as an evolution arising from the interaction of the system ($S$) with an environment ($E$) which form together a closed quantum system as shown in \fig{fig:time_evo}. As a consequence, the overall evolution is unitary and the map $\map$ is given by
\BE
\map(\rho_S)=\Tr_E(U \rho_{SE} U^\dagger)
\EE
where $\rho_{SE}$ describes the initial state of the combined system (see \fig{fig:time_evo}). The resulting map $\map$ is completely positive if the initial state is uncorrelated, that is $\rho_{SE}=\rho_S\otimes\rho_E$.

Every linear map $\mathcal{E}: \mathcal{H}^d \rightarrow \mathcal{H}^d$ is completely characterized by the so called Choi-matrix \cite{Choi1975, Jiang2013}
\BE
\rho_\map\equiv \left(\map_A\otimes \mathds{1}_B\right)\ket{\Phi^+}_{AB}\bra{\Phi^+}
\EE
with $\ket{\Phi^+}=\sum_{j=1}^d\ket{jj}_{AB}$.  The evolved state $\rho_S'$ is then given by the projection of system B onto the transpose of the initial state $\rho_S^T$ that is $\rho_S'=\Tr_B[(\mathds{1}_A\otimes\rho_S^T)  \rho_\map]$. The linear map $\map$ is completely positive, if $\rho_\map$ is positive semidefinite \cite{Choi1975}. Furthermore, the representation of the time evolution of the system by the Choi-matrix allows us to transfer results from state tomography to process tomography by using the Choi-Jamiolkowski isomorphism \cite{Choi1975}.

\subsection{Process tomography}

An unknown linear map can be experimentally determined via process tomography. Due to the Choi-Jamiolkowski isomorphism \cite{Choi1975, Jiang2013}, process tomography corresponds to state tomography of the Choi-matrix \cite{Ziman2008}. In this case, process tomography of a linear map $\map$ acting on a $d$-dimensional system $A$ can be performed with the following steps:
\begin{enumerate}
 \item Provide a $d$-dimensional ancilla system $B$.
 \item Prepare the state $\ket{\Phi^+}_{AB}/\sqrt{d}$.
 \item Prepare the state $\rho_\map$ by applying $(\map_A \otimes \mathds{1}_B )$ on the state  $\ket{\Phi^+}_{AB}$.
 \item Perform state tomography on the resulting state $\rho_\map$ by projecting the state onto the operator basis $M_k \otimes M_j$. 
\end{enumerate}

The basis $\lbrace M_k\rbrace$ of the operator space of $\mathcal{H}_A$ is chosen by projectors of different measurement settings and their measurement outcomes, e.g. Pauli-measurements in $x$-, $y$- and $z$- direction for a single quibt  with the outcomes $\pm 1$ (see e.g. \App{app:qubitPT}). Therefore, it is convenient to relabel the basis $\lbrace M_k\rbrace$ by $M_k^s$ where $s$ denotes the different settings  and $k$ denotes the outcomes. As a consequence, $M_k^s\otimes M_j^r $ represents a complete basis of  $\mathcal{H}_A \otimes \mathcal{H}_B $ with $r,j$ labeling the settings and outcomes of system $B$. As a result, the probability to get the outcome $(j,k)$ for the measurement setting $(r,s)$ is given by
\BE
p_{j,k}^{r,s}=\Tr\left(\rho_\map M_j^r \otimes M_k^s\right)
\EE
with $\sum_{j,k} p_{j,k}^{r,s}=1$ if $(j,k)$ contain all possible outcomes. The Choi-matrix $\rho_\map$ can then be reconstructed via
\BE
\rho_\map = \sum\limits_{j,k,r,s} p_{j,k}^{r,s} D_j^r\otimes D_k^s
\EE
where $\lbrace D_n \rbrace$ forms the dual basis of $\lbrace M_m\rbrace$ with $\Tr(D_nM_m)=\delta_{n,m}.$  

The quantum process tomography scheme  described  above exhibits the experimental problem, that a physical system twice as big as the system, on which the map acts, needs to be available and controllable. This is often not the case. Nevertheless, the scheme  for process tomography described  above can  also be applied without an additional ancilla system, as is outlined in the next paragraph. 

The  expectation value of the observable $\hat A \otimes \hat B$ of the state $\ket{\Phi^+}_{AB}\bra{\Phi^+}$ is equivalent to
\BE
\Tr\left[\hat A \otimes \hat B\ket{\Phi^+}_{AB}\bra{\Phi^+} \right]= \sum\limits_j b_j \Tr\left[\hat A \ket{b_j^\ast}_A\bra{b_j^\ast}\right]
\EE
with $b_j$ being the eigenvalues of $\hat B$ and $\ket{b_j^\ast}$ being the complex conjugate of the corresponding eigenvectors (see e.g. \cite{Steffen2012}).
As a consequence, instead of performing the above described quantum process tomography scheme with ancilla system, (i) we prepare the system in different basis states $M_k^{\ast}$ corresponding to the measurement outcome $M_k$ for the measurement on system $B$, (ii) apply the map $\map$ on the system and finally (iii) perform the measurement $M_j$ on the system.

An important difference between the two schemes is that, in the first case, the map on system $A$ is first applied before we define the initial state of the system by the projection of system $B$. In the second case, the projection of system $B$ is equivalent to the preparation process, which is performed before the application of the map. If the initial state of the environment is independent of the system, then the preparation/projection of system $B$ and the map $\map$ do commute. If the preparation of the system induces correlations with the environment, then the map $\map$  may depend on the preparation process, and therefore they do not commute. In this case, the description of the time evolution of the system by $\map$ is incomplete since it does not involve the preparation process.

\subsection{Time evolution with initial system-environment correlations }

Process tomography in the presence of system-environment correlations can lead to reconstructed Choi-matrices with negative eigenvalues \cite{Pechukas1994, Wood2009,Modi2010,Ziman2006}. However, this does not mean that the underlying process itself is non CP; it only indicates that the chosen description is incorrect \cite{Schmid2018}. So far, there exists no general theoretical framework to describe arbitrary (non CP) maps. Theoretical frameworks of non CP maps exist only for restricted subsets of correlations and/or a restricted subset of states \cite{Carteret2008, Vacchini2016, Ringbauer2015} and are an important research topic until today. For example, the time evolution can be described by a superchannel, taking the preparation procedure as input, if the joined system-environment state $\rho_{SE}$ is initially correlated/entangled and the preparation procedure only acts on the system. \cite{Ringbauer2015}. Yet, arbitrary mappings $\rho_S\rightarrow \rho_s'$ are possible if infinite system-environment correlations are allowed \cite{Ziman2006,Carteret2008}. Assuming, for example, that the environment consists of an infinite number of copies of $\rho_S$,  the environment can gain perfect knowledge about the system state $\rho_S$. Consecutively, it can prepare $\rho_S'$ and transfer this state via the SWAP operation into the system. 
Therefore, a prediction of the time evolution of $\rho_S$ without any knowledge about the state of the environment is impossible due to a lack of information.

This incompleteness of $\map$ as description of the time evolution of $\rho_S$ can be illustrated with the following example: let us assume that the preparation of the system represented by a single qubit prepares the environment, also represented by a single qubit, in exactly the same state, that is $\rho_E=\rho_S$. This is a pure classical correlation, it does not involve any quantum correlations. The time evolution of the joined system is given by
\BE
U=\exp[\I \frac{\pi}{4}\;\hat{z}_S\hat{z}_E]\exp[\I \frac{\pi}{4}\;\hat{x}_S] \label{eq:time_evo}
\EE
with $\hat z$ and $\hat x$ denoting the Pauli matrices.

A complete basis $\lbrace M_j \rbrace$ of the system is given by the eigenstates $\ket{0},\ket{1}$ of $\hat z$ and $\ket{+}$ and $\ket{\I}$ denoting the eigenstates of $\hat x$ and $\hat y$ with eigenvalue $+1$. As a consequence, the time evolution of these states, given by
\BE
\begin{array}{cccccc}
\ket{0}\bra{0}&\rightarrow &\ket{+}\bra{+},\quad&\ket{1}\bra{1}&\rightarrow& \ket{+}\bra{+}\\
\ket{+}\bra{+}&\rightarrow &\mathds{1}/2,\quad&\ket{\I}\bra{\I}&\rightarrow &\ket{1}\bra{1},
\end{array}\label{eq:basis_evo}
\EE
completely describes the map $\map$. However,  no oracle can perform such a time evolution from knowing $\rho_S$ alone without additional information, since it is impossible to distinguish the input state with only a single copy. This time evolution is only possible with additional information given here by the copy of the state provided by the environment. In this way, the  time evolution described here  is not linear anymore. As a consequence, the attempt to describe this time evolution with the help of a liner map, given by the  resulting Choi-matrix
\BE
\rho_\map = \frac{1}{2}\left(\begin{array}{cccc}
	1&1&-\I&-1-\I\\ 1&1&-1-\I&\I \\ +\I&-1+\I&1&1 \\ -1+\I & -\I & 1 & 1
	\end{array}\right)\label{eq:Choi_theo}
	\EE
leads to a non-physical result given by the negative eigenvalue $\lambda_\map=-\sqrt{3}/2$. This does not mean that the states $\rho_S'$ of the system after the time evolution are non-positive (see \eq{eq:basis_evo}). The map describing this time evolution is still positive. However, it is not completely positive. This means, if our system $S$ is coupled to another system $R$, then the time evolution according to $\map_S$ may lead to  a non-positive state $\rho_{RS}'=(\mathds{1}_R\otimes \map_S)(\rho_{RS})$. This is due to the incomplete description given by $\map$ of our system. Without the exact definition of the preparation process of $\rho_{RS}$ and its effects on the environment, we cannot predict the time evolution of the composite system. 

Let us assume, for example, that the environmental qubit is only affected by single qubit rotations acting on the system qubit, but not by the interaction between the systems $S$ and $R$. Then, a state such as $(\ket{00}_{RS}+\ket{11}_{RS})/\sqrt{2}$ can be prepared with a CNOT-gate with $S$ being either the control or the target qubit leading to two different  initial states
\begin{eqnarray}
\ket{\psi_1}_{RSE}&=&\frac{1}{\sqrt{2}}(\ket{00}_{RS}+\ket{11}_{RS})\ket{0}_E, \\
\ket{\psi_2}_{RSE}&=&\frac{1}{\sqrt{2}}(\ket{00}_{RS}+\ket{11}_{RS})\ket{+}_E .
\end{eqnarray}	
As a consequence, the description of the time evolution of the system by $\map$ is incomplete and may lead to non-physical predictions expressed by the non-completely positivity of the map $\map$.

\section{Consistency test\label{sec:test}}

As discussed in the introduction, there exist different reasons for the appearance of negative eigenvalues in experimentally reconstructed processes. In what follows, we describe a method that tests if the observed negativity might be the result of statistical effects, or if the assumed model underlying the reconstruction process should be revisited.
Our method for detecting systematic errors in quantum process tomography is based on a witness test, similar to an entanglement witness, and is based on certification of experimental errors in state tomography \cite{Moroder2013, Knips2015, Schwemmer2015}. That is, we construct an observable $Z_w=\ket{\lambda}\bra{\lambda}$, called witness, which is positive semidefinite for the assumed model. Therefore, the appearance of negative average values with sufficient significance indicates an inconsistency with the assumed model. The significance can be tested with the help of the Hoeffding inequality \cite{Hoeffding1963}.

Now, we first analyse quantum process tomography with the help of an ancilla system. In this case, the results of Ref.\cite{Moroder2013} about certifying experimental errors in state tomography can be directly applied. Then, we use the Choi-Jamiolkowski isomorphism \cite{Choi1975, Jiang2013} to translate this test to the more commonly used form of  quantum process tomography without an ancilla system.

The expectation value $\langle Z_w \rangle=\bra{\lambda} \rho_\map \ket{\lambda}$ given by the projection of $\rho_\map$ on an arbitrary state $\lambda$ must be positive, if $\rho_\map$ is positive semidefinite. To evaluate $\langle Z_w\rangle$ we expand $Z_w$ into the basis $\lbrace M_k \otimes M_j \rbrace$ (see \Sec{sec:ncp_maps}).
The basis $\lbrace M_k\rbrace$ of the operator space of $\mathcal{H}_A$ is chosen by projectors of different measurement settings and their measurement outcomes. Measurement outcomes of the same measurement setting are not independent of each other because they sum up to one. Therefore, we relabel the basis $\lbrace M_k\rbrace$ by $M_k^s$ where $s$ denotes the different settings  and $k$ the outcomes (see e.g. \App{app:qubitPT}). In general, not all measurement outcomes are necessary to obtain a complete basis, for example,  we do not use the eigenstates corresponding to the eigenvalue $-1$ of the Pauli $x$ ($M_-^x=\ket{-}\bra{-}$)  and of the Pauli $y$ matrix ($M_-^y=\ket{-\I}\bra{-\I}$) for single qubit process tomography. Furthermore, we assume that each measurement setting $(r,s)$ is used $N_{RS}$ times.
As a result, the witness $Z_w$ can be expanded by
\BE
Z_w \equiv \sum_{(r,j)(s,k)}w_{j,k}^{r,s} M_j^r \otimes M_k^s\label{eq:witness},
\EE
where $w_{j,k}^{r,s}=0$ for operators which are not part of the basis chosen.
With the help of this expansion, we are able to determine the expectation value
\BE
\langle Z_w \rangle = \sum_{(r,j)(s,k)}w_{j,k}^{r,s} f_{j,k}^{r,s}
\EE
where $f_{j,k}^{r,s}$ denotes the observed frequencies to get the result $(j,k)$ for the measurement setting $(r,s)$. If these frequencies are the result of a quantum model, then the probability $P$ to get a negative expectation value $\Tr[Z_w\rho_\map]=w\cdot f< -t$ for  $t>0$ is bounded by \cite{Moroder2013}
\BE
\textrm{Prob}\left[w \cdot f \leq -t\right] \leq \exp\left[-2\frac{t^2N_{RS}}{\sum_{r,s}(w_\textrm{max}^{r,s}-w_\textrm{min}^{r,s})^2}\right]\label{eq:Hoeffding}
\EE
which follows from the Hoeffding inequality \cite{Hoeffding1963} (see also \App{app:Hoeffding}). Here $w_\textrm{max}^{r,s}$ and $w_\textrm{min}^{r,s}$ denote the maximal and minimal expansion coefficients for the measurement setting $(r,s)$.   If this probability is very low and lies below a predefined threshold $\alpha$ (common values are $5\%$ or $1\%$ \cite{Knight}), then the consistency test fails. In this case, the assumed model is very unlikely and the experiment should be revisited. In summary, the consistency test
consists of three steps:
\begin{itemize}
\item Choose a witness $Z_w=\ket{\lambda}\bra{\lambda}$ (see \Sec{sec:examples}).
\item Expand $Z_w$ into the basis $ M_j^r \otimes M_k^s$.
\item If $\langle Z_w \rangle <0$ then determine the probability $P$ and compared it to the predefined threshold.
\end{itemize}

In the case of process tomography without ancilla system,  the expectation value of the witness $Z_w$ is given in a similar way by
\BE
\Tr[Z_w \rho_\map]= \frac{1}{d}\sum w_{j,k}^{r,s} \Tr\left[M_k^s \map\left(M_j^{r\ast}\right)\right].
\EE
Here, the system was prepared in the state $M_k^{s\ast}$, evolved in time, and measured in the basis $M_j^r $. The witness $Z_w$ determined by the coefficients $w_{j,k}^{r,s}$ stays the same. Only the way in which the frequencies $f$ are evaluated is different. For process tomography with ancilla system, the frequency $f_{j,k}^{r,s}=f_j^r \cdot f_s^k$ is the product of the observed frequencies of both systems and the probability to obtain outcome $k$ for system $B$ is equally distributed independent of the setting $s$. In the other case, $f_{j,k}^{r,s}=f_j^r \cdot p_s^k$ where $p_s^k$ is the probability that we prepare system A in the state $(M_k^s)^\ast$. In general, only the states $(M_k^s)^\ast$ necessary to obtain a complete basis are prepared. Therefore, $p_s^k$ can be zero for some states. However, the Hoeffding inequality does not depend on the exact probability distribution; only on the boundaries which are the same for both cases. Therefore, the consistency test for process tomography stays always the same no matter how we perform the process tomography.

The witness $Z_w$ depends on the map $\map$. However, it is important  not to use the same data to determine the witness $Z_w$ and to perform the witness test. If we scan a large set of data for any correlation, we will always find a correlation with high significance due to statistical fluctuations, see e.g. \cite{Hjortrup2016}. Therefore, the witness $Z_w$ should be determined by a different set of data or by testing theoretically assumed errors as we will demonstrate in the next section.


\section{Examples\label{sec:examples}}

To investigate the potential of the scheme described above for  discriminating between statistical and systematic errors, we simulate and experimentally perform  process tomography of several single-qubit quantum channels.

For each simulation/experiment, we first prepare $N_{RS}$ of each of the states $M_k^{s\ast}$ followed by a measurement described by $M^r_j$. Then, we reconstruct the state $\rho_\map$ (see \App{app:qubitPT}). To determine the best witness $Z_w$ we use different methods. For the simulations, we divide the data set into two parts. The first part is used to determine $Z_w$, with the second part we perform the consistency test. Another option is to guess the underlying error. In this case, a Choi-matrix $\rho_\map^\text{theo}$ including the assumed error can be theoretically calculated. With the help of $\rho_\map^\text{theo}$ the witness $Z_w$ can then be predicted. We have used this procedure to test our experimentally generated data.

\subsection{Simulation\label{sec:simulation}}

\begin{figure}[t]
\includegraphics[width=0.4\textwidth]{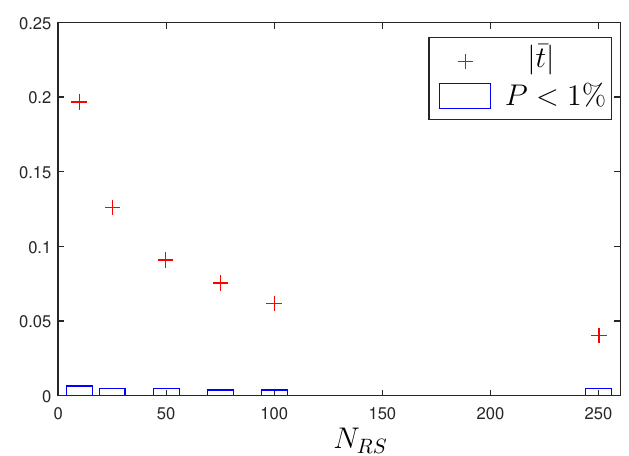}\\
\caption{Average negative expectation value $|\bar t|$ and proportion of results with probability $P<1\%$ for the simulation of $N=10^4$ process tomographies, for different measurement repetitions $N_{RS}$ in the presence of only statistical errors. The time evolution is given by \eq{eq:time_evo_nc}. }
\label{fig:Rx}
\end{figure}

\begin{figure}[t]
\includegraphics[width=0.4\textwidth]{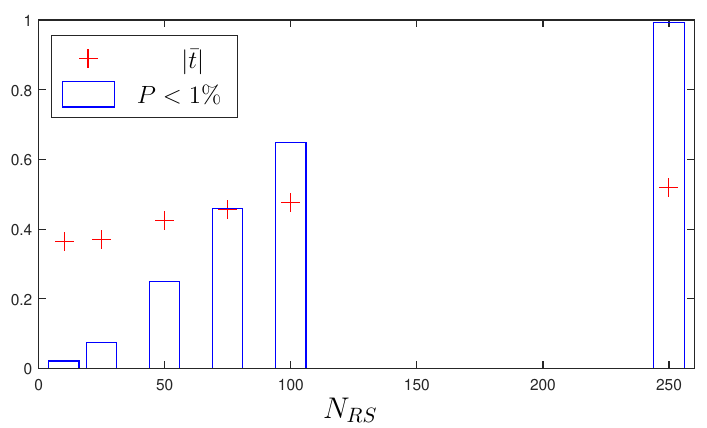}\\
\caption{ Average negative expectation value $|\bar t|$ and proportion of results with probability $P<1\%$ for the simulation of $N=10^4$ process tomographies, for different measurement repetitions $N_{RS}$ in the presence of a detuning $\delta=0.25\Omega$ of the radio frequency (RF) pulses performing single qubit rotations. Here, $\Omega$ denotes the Rabi frequency of the qubit transition. The time evolution is given by \eq{eq:time_evo_nc}.}
\label{fig:detuned}
\end{figure}

\begin{figure}[t]
\includegraphics[width=0.4\textwidth]{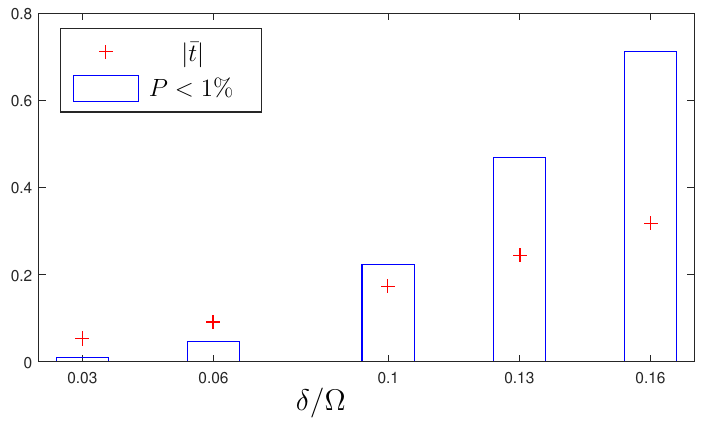}\\
\caption{ Proportion of results with probability $P<1\%$ for the simulation of $N=10^4$ process tomographies and for $N_{RS}=250$ measurement repetitions for different detunings $\delta$.}
\label{fig:detuned_omega}
\end{figure}

\begin{figure}[t]
\includegraphics[width=0.4\textwidth]{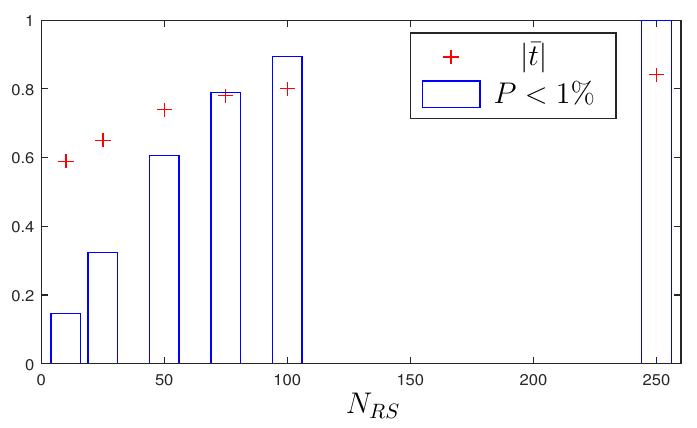}\\
\caption{ Average negative expectation value $|\bar t|$ and proportion of results with probability $P<1\%$ for the simulation of $N=10^4$ process tomographies, for different  measurement repetitions $N_{RS}$ in the presence of initial correlations between the system and its environment.  The time evolution is given by \eq{eq:time_evo}.}
\label{fig:Uzz}
\end{figure}

 If the state $\rho^{(1)}_\map$ possesses negative eigenvalues, then the best witness $Z_w=\ket{\lambda}\bra{\lambda}$ is given by the eigenstate \ket{\lambda_\text{min}} of $\rho^{(1)}_\map$ with the most negative eigenvalue $\lambda$. In general, every witness $Z_w$ with $\Tr[Z_w \rho_\map]=-t_Z<0$ can be used for the consistency test. However, the smaller $|t|$, the greater the probability $P$ to obtain the same negativity due to statistical results. If $\text{Prob}[w\cdot f \leq -t_Z]$ is larger as the chosen significance level $\alpha$, it is not possible to obtain meaningful results. Yet, increasing the number $N_{RS}$ of measurements can help.

We determine the coefficient $w_{j,k}^{r,s}$ by representing $Z_w$ as a sum over all $M_j^r\otimes M_k^s$ (see \eq{eq:witness}), and evaluate $C=\sum_{r,s}(w_\textrm{max}^{r,s}-w_\textrm{min}^{r,s})^2$.

Afterwards, a second round with $N_{RS}$ preparations and measurements of each setting is performed, which leads to $\rho^{(2)}_\map$. Finally, we estimate the average value $\Tr[Z_w\rho^{(2)}_\map]=-t$ and the corresponding probability $\textrm{Prob}[w\cdot f \leq -t]$. This probability is also called p-value in hypothesis testing \cite{Knight}. The p-value is an upper bound for the probability that the observed average value was generated by the assumed statistical model. Commonly, the hypothesis is discarded if the p-value is smaller than $\alpha=5\%$ or $\alpha=1\%$. In this case, we assume the observed discrepancy between the theoretically predicted expectation value and the observed average value is not only caused by statistical errors but by systematic errors.

In what follows, we simulate the process tomography of three different processes: (i) a perfect process tomography with only statistical errors, (ii) a process tomography with wrong preparation and measurement directions  (iii) a process tomography with initial correlation between the system and the environment.

For the first two cases, the time evolution is given by a single rotation around the $x$- asis
\BE
\map(\rho)= \E^{\I\pi \hat{x}/4} \rho \E^{-\I\pi \hat{x}/4}\label{eq:time_evo_nc}
\EE
that is equivalent to \eq{eq:time_evo} without system-environment interaction.

In the second case, we assume an experiment with trapped ions. Here, the preparations and measurements in $x$- and $y$- directions are performed by applying additional $\pi/2$ rotations around the $y$- or $x$- axis to the ions, followed by a measurement in the $z$-direction. A typical error in trapped-ion experiments is detuning. If the RF field , used for the $\pi/2$ rotations, is  detuned by $\delta$ from the qubit resonance, then the preparation and the measurement directions are not perfect anymore. The detuning will lead to a different rotation angle $\theta' = \theta \cdot \sqrt{\Omega^2+\delta^2}/\Omega$, with $\Omega$ being the Rabi frequency, and the rotation axis $\vec{n}$ will be tilted towards the $z$-axis with $\vec{n}\cdot\vec{e_z}=\delta/\sqrt{\Omega^2+\delta^2}$.

In the third case, we assume that another ion is sitting in the trap playing the role of the environment. We simulate a strong initial correlation between system and environment by preparing the environment in the same state as the system. This is a simplified version of the case, where the environment has perfect knowledge about the system state and arbitrary mappings $\rho_S \rightarrow \rho_S'$ are possible. The time evolution is given by \eq{eq:time_evo}.

We simulated the above described cases with the help of MATLAB. In \fig{fig:Rx}-\ref{fig:Uzz} we summarize the percentage of process tomographies with $\textrm{Prob}[w\cdot f<-t]<0.01$ as well as the average negative expectation value $-t=\Tr[Z_w\rho_\map]$  for $10^4$ simulated tomographies. Each tomography corresponds to the reconstruction of the process matrix $\rho_\map$ from 12 different measurement settings, each repeated $N_{RS}$ times. 

If only errors are present that fall into category (i), statistical errors,  then the percentage of discarded tomographies stays the same independent of the number of measurements per setting as shown in \fig{fig:Rx}. Here, we can observe the general behavior of $P$ and $t$ which is independent of the absolute value of the chosen significance level $\alpha$ and typical for all time evolutions where the Choi-matrix possesses eigenvalues equal to zero. The number of discarded events is independent of the number $N_{RS}$ of measurements. Only the amount of discarded events depends on $\alpha$ and is typically smaller than $\alpha$ because the direction of the measurement with the smallest eigenvalue also randomly changes. The probability to observe a negative average value $\langle Z_w\rangle$ is approximately $50\%$ \cite{Knips2015} independent of $N_{RS}$, Yet  the average negativity of $t$ decreases with $N_{RS}$. 

However, if systematic errors are also present, such as in categories  (ii) and (iii), the more measurements we perform, the more tomographies we reject. For example about $N_{RS}=250$ measurements per setting are necessary to detect a detuning of $\delta/\Omega$=0.25 reliably as demonstrated in \fig{fig:detuned}. For this case, the average negative value $|t|$ increases for small $N_{RS}$ until it reaches its true value $|t|=|\lambda_\text{min}|\approx 0.56$. This effect results from statistical fluctuations of the direction of the eigenstate $\ket{\lambda_\text{min}}$ for small $N_{RS}$. In general, the mean negative value $|t|$ and the number of measurements per setting $N_{RS}$ necessary to reliably detect a systematic error depend  on the magnitude of the systematic error. The larger the systematic error, e.g. the detuning $\delta$ in case (ii), the fewer measurements we need to detect it as displayed in \fig{fig:detuned_omega}.

The behavior of our consistency test in the presence of (iii) initial correlations between system and environment, as shown in \fig{fig:Uzz}, is similar to the behavior for case (ii) since both errors are systematic errors. However, for the example in case (iii) we get higher discarding rates than in case (ii) since the minimal eigenvalue $\lambda_\text{min}\approx -0.87$ for our example in case (iii) is smaller than the one for case (ii).

\subsection{Experimental results\label{sec:experiment}}

In the following we describe the experimental implementation of a process tomography with initial correlation between the system and its environment. Both the system and the environment are represented by a single qubit, each realized by a single trapped $^{171}\mbox{Yb}^+$ ion. They form a Coulomb crystal exposed to a static magnetic field gradient of 19 T/m in a linear Paul trap with an axial trap frequency of $2\pi \times 120$ kHz and a radial trap frequency of $2\pi \times 590$ kHz. The state $\ket{0}$ is represented by the energy level $\ket {^2S_{1/2},F=0}$ and $\ket{1}$ by $\ket{^2S_{1/2},F=1,m_F=+1}$ \cite{Khromova2012,Woelk2015,Piltz2016}.

The vibrational excitation is reduced by Doppler cooling followed by RF sideband cooling and is characterized by the mean vibrational quantum number of the center-of-mass mode $\langle n \rangle<15$ \cite{Sriarunothai2018}. Then, the qubits are initialized in the state $\ket{0}$ by optical pumping. Single-qubit rotations with the help of RF pulses near 12.6 GHz corresponding to the respective qubit transitions are performed to prepare the system qubit and the environmental qubit in the same desired initial states given by  $\ket{0}, \ket{1}, \ket{+}$, and $\ket{i}$. The time evolution (see \eq{eq:time_evo}) is realized with the help of MAgnetic Gradient Induced Coupling (MAGIC) \cite{Wunderlich2002,Khromova2012,Piltz2016, Woelk2017}. The evolution time takes 5.8 ms matching the J-coupling between 2 ions.
The qubit dephasing is protected by dynamical decoupling (DD) pulses \cite{Piltz2013} applied to both qubits using the Universally Robust (UR) DD sequence \cite{Genov2016}.
These DD-pulses are applied during the evolution time (for more epxerimental details see \App{app:ExpSeq}). Finally, the measurement on the system qubit is performed in different bases ($\sigma_x, \sigma_y, \sigma_z$) with the help of single qubit rotations and by detecting state selectively scattered resonance fluorescence using an electron multiplying charge coupled device (EMCCD). Detailed information about the experimental setup is available elsewhere \cite{Khromova2012,Woelk2015,Piltz2016}.

Each preparation and measurement setting was repeated $N_{RS}=394$. The resulting experimental reconstructed Choi-matrix is given by
\begin{widetext}
\BE
\rho_\map^{(\text{exp})}=
\frac{1}{2}\left(\begin{array}{rrrr}
   0.99+0.00\I  & 0.87+0.11\I &  0.10-0.83\I  &-0.89-0.74\I\\
   0.87 - 0.11\I &  1.01 + 0.00\I & -1.04 - 1.09\I&-0.10 + 0.83\I\\
   0.10 + 0.83\I&-1.04 + 1.09\I&0.82 + 0.00\I&0.84 - 0.22\I\\
  -0.89 + 0.74\I & -0.10 - 0.83\I&   0.84 + 0.22\I&   1.18 + 0.00\I
\end{array}\right),
\EE
\end{widetext}
with statistical error of $\Delta\rho_{j,k}=\pm 0.025$ and a minimal eigenvalue of $\lambda_\text{exp}=-0.70$. We used the eigenstate $\ket{\lambda_\text{theo}}$ corresponding to the eigenvalue $\lambda_\text{theo}=-\sqrt{3}/2\approx -0.87$ of theoretically predicted Choi-matrix \eq{eq:Choi_theo} to determine the witness $Z_w=\ket{\lambda_\text{theo}}\bra{\lambda_\text{theo}}$.
The resulting expectation value $\Tr[Z_w \rho_\map^{(\text{exp})} ]=-0.67$ is with a probability of $P<4 \cdot 10^{-20}$ the result of purely statistical effects. As a consequence, our consistency test revealed the error of the experimentally realized process with the help of the theoretically predicted witness $Z_w$. On the other hand, the theoretically predicted witness for just simple detuning of $\delta=0.25\Omega,0.5\Omega$ or $\Omega$ does not reveal any inconsistencies.

In general, our consistency test only makes a statement about whether the assumed model is consistent with the observed data, and whether the negativity we observe is severe or not. In this sense, it can only falsify a model, but never verify it. The test itself, especially if the witness is reconstructed via a first set of data, makes no statement about the systematic error itself. To obtain information about the sort of error, we have to study the influence of different possible error sources on the data. Here, it is also helpful to not only have a look on the Choi-matrix itself, but also on the reconstruction of the time evolution of test states $\rho_j'$ which can be extracted from the same data.

In \App{sec:rec_states} we summarize the reconstructed states $\rho_j'$ for the experimental data as well as for other assumed errors such as detuned RF pulses applied to the qubits. Detuned pulses lead to wrong preparation of the inital test states $\rho_j$, a different time evolution, and wrong measurement directions. Wrong preparation and measurement directions can lead to negative eigenvalues of the Choi-matrix as well as negative eigenvalues of the reconstructed states. On the other hand,  initial correlation between the system and the environment leads only to negative eigenvalues of the Choi-matrix.

Other errors, such as detection errors, dephasing, and spontaneous decay change the time evolution but will lead to a proper Choi-matrix with positive eigenvalues. However, they can explain the difference between the theoretically predicted Choi-matrix for our specially designed correlation and the observed experimental data. The purity of the reconstructed state $\rho_j'$ with $j=1,2,4$ is very high (see \App{sec:rec_states}). Therefore, we assume that dephasing and decay do not play an important role in our experiment.

A detection error $\varepsilon$ will shift extreme expectation values such as $\langle \hat{z}\rangle=\pm 1$ towards the average $\langle \hat{z}\rangle=0$. If the detection errors for the two eigenstates are different, the average $\langle \hat{z}\rangle=0$ will be additionally shifted towards the measurement value with smaller error. These are the so-called dark states $\ket{0},\ket{-}$ and $\ket{-\I}$ in our case which lead to reduced average values.  However, this behavior can only be observed in some of our measurements whereas the average values are shifted sometimes also in the other directions (see \App{sec:rec_states}).  This could be the result of stray light from the fluorescence laser which leads to population trapping in the states $\ket{^2S_{1/2},F=1,m_F=0}$ and $\ket{^2S_{1/2},F=1,m_F=-1}$. This leads together with DD to increased average values.

 The overall effect of these three possible errors (detuned pulses, asymmetric detection error and stray light) on the time evolution of the test states can be seen in \App{sec:rec_states} and fits very well the experimental data.

Another error source in process tomography are drifts. These errors can be treated in the same way as in state tomography, e.g. they can be decreased by randomly switching  between different measurements setups. Another method is to describe the observed data with the help of additional parameters and use the Akaike Information Criterion to judge if this model leads to a better description of the system \cite{Enk2013, Guta2012}.

\section{Conclusion \label{sec:conclusion}}
In this paper, we discuss and demonstrate, via experiments and simulations,  how non-completely positive maps can appear in quantum process tomography. Furthermore, we introduced a simple tool for data analysis to discriminate between statistical and systematic errors. Using this tool, initial correlations between the system and its environment are experimentally detected with less than 400 repetitions for each measurement setup. Furthermore, the witness constructed specifically to identify systematic errors in the preparation and measurement process (caused by detuned RF pulses) did not detect errors caused by initial correlations. This suggests that the witness test can not only discriminate between systematic and statistical errors but also between preparation/measurement errors and correlations. However, further studies on the different influence of these different errors on the Choi-matrix need to be carried out  to confirm this conjecture.  The consistency tests introduced here for data collected in the course of quantum process tomography can be carried out with small additional effort compared to collecting the experimental data and bring significant insights.

In general, the  consistency test introduce here can not only be applied to full process tomographies, but also to incomplete measurements. In this case, our test will be  sensitive solely to some systematic errors. In general, all witnesses with $\Tr[Z_w\rho_\map]< 0$ can be used for our hypothesis test. However, the test is more sensitive, the smaller $\Tr[Z_w\rho_\map]$. Therefore, it might be advantageous to determine possible test states and measurements via a first full process tomography to identify possible problems of an experiment such as drifting laser- or  radio frequencies. Later on, only the determined test state needs to be prepared and appropriate measurements need to be performed to observe the thus identified problem and to appropriately counteract while carrying out experiments.

If the test proposed here suggests an inconsistency, it is also possible to directly test for initial system-environment correlations by using a witness based on distinguishability  \cite{Laine2010,Gessner2011}, or purity \cite{Rossatto2011}. Such tests would require additional data collection and are beyond the purpose of this paper. Another possibility is to characterize the reduced dynamics in the presence of initial system-environment correlations, if the preparation procedure only acts on the system \cite{Ringbauer2015}. Yet, this characterization  cannot be applied to the experimental example presented in section IV, because the preparation procedure used there also acts on the qubit representing the environment.

\section*{Acknowledgement}
 S.W. thanks M. Kleinmann for fruitful discussions. We acknowledge funding from Deutsche Forschungsgemeinschaft.
G.S.G. acknowledges support from the European Commission's Horizon 2020 research and innovation program under Marie Sk\l{}odowska-Curie grant agreement number 657261.

\begin{appendix}
\section{Single qubit process tomography\label{app:qubitPT}}

For a single qubit, a possible measurement basis $M_k^s$ is given by
\begin{eqnarray}
M^z_0=\ket{0}\bra{0}&&M^z_1=\ket{1}\bra{1}\nonumber\\
M^x_1=\ket{+}\bra{+}&&M^y_1=\ket{\I}\bra{\I}
\end{eqnarray}
with $\ket{+}=(\ket{0}+\ket{1})/\sqrt{2}$ and $\ket{\I}=(\ket{0}+\I\ket{1})/\sqrt{2}$. The reconstruction of the state $\rho_\map$ is given by
\BE
\rho_\map= \sum\limits_{(r,j),(s,k)}p_{j,k}^{r,s} \;D_j^r \otimes D_k^s
\EE
with the probabilities $p_{j,k}^{r,s}= \Tr[M_j^r \otimes M_k^s \rho_\map]$ and the Dual-basis
\begin{eqnarray}
D^z_0=\frac{1}{2}\left(\begin{array}{cc}
2 & -1+\I \\ -1-\I&0
\end{array}\right) && D^z_1=\frac{1}{2}\left(\begin{array}{cc}
0 & -1+\I \\ -1-\I&2
\end{array}\right)\nonumber\\
D^x_0=\left(\begin{array}{cc}
0 & 1 \\ 1&0
\end{array}\right) && D^y_0=\left(\begin{array}{cc}
0 & -\I \\ \I&0
\end{array}\right)
\end{eqnarray}


\section{Hoeffding inequality\label{app:Hoeffding}}
In this appendix, we shortly summarize the Hoeffding inequality and the results of Ref. \cite{Moroder2013}.
In statistics, the observed sum of bounded independent random variables $\bar{X}=\sum_\ell^L X_\ell$  may vary from its expectation value $E[X ]$ due to limited sample size. The probability that they vary from each other by more than $t$ is upper bounded by \cite{Hoeffding}
\BE
\text{Prob}[E[ X]-\bar{X}  \geq t]\leq \exp\left[-\frac{2t^2}{\sum_{\ell=1}^L (b_\ell-a_\ell)^2}\right]\label{eq:B1}
\EE  
if $a_\ell \leq X_\ell \leq b_\ell$. The random variables for the consistency test described in \Sec{sec:test} are given by
\BE
X_\ell = \frac{1}{N_{RS}} \sum\limits_{(r,j)(s,k)}w_{j,k}^{r,s}\;n_{j,k}^{r,s}(\ell)
\EE
where $n_{j,k}^{r,s}(\ell)=1$ if we have used the setting $(s,r)$ in the $\ell$-experiment and obtained the result $(j,k)$. As a consequence, $X_\ell$ is bounded by $w_\text{min}^{r(\ell),s(\ell)}/N_{RS}\leq X_\ell \leq w_\text{max}^{r(\ell),s(\ell)}/N_{RS}$.  The right side of \eq{eq:B1} can be rewritten as
\BE
\exp\left[-\frac{2t^2}{\sum_{\ell=1}^L (b_\ell-a_\ell)^2}\right] = \exp\left[-2\frac{t^2N_{RS}}{\sum_{r,s}(w_\textrm{max}^{r,s}-w_\textrm{min}^{r,s})^2}\right]
\EE 
where we sum  only over all settings $(r,s)$ instead of all variables $\ell$. Note, that we use each setting $N_{RS}$ times. The condition $w\cdot f= \bar{X}\leq -t$ on the left side of \eq{eq:Hoeffding} is equivalent to
\BE
\bar{X}\leq -t \Leftrightarrow E(X)-\bar{X}\geq t+E(X). 
\EE
The probability for this is upper bounded according to \eq{eq:B1} by
\BE
\text{Prob}[E[ X]-\bar{X}  \geq \tilde{t}]\leq \exp\left[-2\frac{\tilde{t}^2N_{RS}}{\sum_{r,s}(w_\textrm{max}^{r,s}-w_\textrm{min}^{r,s})^2}\right] 
\EE
where we have defined $\tilde{t}=E(X)+t$. Note, $\tilde{t}\geq t$ since $E[X]>0$ and thus we finally arrive at \eq{eq:Hoeffding}.


\section{Experimental sequence \label{app:ExpSeq}}
The experimental sequence is shown in \tab{tab:pulse_sequence}. The system qubit and the environmental qubit are initialized in the state $\ket{00}$. Then, the states $\ket{0}$, $\ket{+}$, $\ket{i}$, and $\ket{1}$ are prepared in step 0 by single qubit rotations given by
\BE
R(\theta,\varphi)=\exp[\I \frac{\theta}{2} (\hat{x}\cos \varphi-\hat{y} \sin \varphi) ].
\EE
Step 1 and the conditional evolution perform a controlled-phase gate, where the environmental qubit is the control qubit and the system qubit is the target qubit. Step 2 to 22 describe the conditional evolution
\BE
U^{(jk)}(t)=\exp[\I \frac{t}{2} J_{jk} \hat{z}_j \otimes \hat{z}_k]
\EE
together with the pulses for dynamical decoupling (DD). Here, $J_{j,k}$ describes the coupling between ion $j$ and ion $k$. For our experiment, we used $N_p=100$ pulses for DD, which means we repeated step 2 to 22 for 10 times. We used a total conditional evolution time $\tau=\pi /(2J_{j,k})=5.8$ ms. Step 23 describes the rotation of the system qubit  necessary to perform spin measurements in $x$-, $y$- or $z$-direction.

\begin{center}

	\begin{table}
	\caption{Pulse sequence for single qubit process tomography. The superscripts (1), (2), and (12) indicate  that an operation is applied to the system, to the environment, or to both qubits, respectively. $\mathds{I}$ represents the identity operator. Each single qubit rotation or RF pulse is specified by a pulse area and phase given within parentheses.}
	\label{tab:pulse_sequence}
	\begin{tabular}{|c | c|}
	\hline
	Number & RF-pulse  \\
	\hline
	0 & $\mathds{I}$ or $R^{(1)}(\frac{\pi}{2},0)$ or $R^{(1)}(\frac{\pi}{2},\frac{\pi}{2})$ or $R^{(1)}(\pi,0)$ \\
	\hline
	0 & $\mathds{I}$ or $R^{(2)}(\frac{\pi}{2},0)$ or $R^{(2)}(\frac{\pi}{2},\frac{\pi}{2})$ or $R^{(2)}(\pi,0)$ \\
	\hline \hline
	1 & $R^{(1)}(\frac{\pi}{2},0)$  \\
	\hline
	\hline
	2 & $U^{(12)}(\frac{\tau}{2N_p})$  \\
	\hline
	3 & $R^{(1)}(\pi,0) \otimes R^{(2)}(\pi,0)$  \\
	\hline
	4 & $U^{(12)}(\frac{\tau}{N_p})$  \\
	\hline
	5 & $R^{(1)}(\pi,\frac{4\pi}{5}) \otimes R^{(2)}(\pi,\frac{4\pi}{5})$  \\
	\hline
	6 & $U^{(12)}(\frac{\tau}{N_p})$  \\
	\hline
	7 & $R^{(1)}(\pi,\frac{2\pi}{5}) \otimes R^{(2)}(\pi,\frac{2\pi}{5})$  \\
	\hline
	8 & $U^{(12)}(\frac{\tau}{N_p})$  \\
	\hline
	9 & $R^{(1)}(\pi,\frac{4\pi}{5}) \otimes R^{(2)}(\pi,\frac{4\pi}{5})$  \\
	\hline
	10 & $U^{(12)}(\frac{\tau}{N_p})$  \\
	\hline
	11 & $R^{(1)}(\pi,0) \otimes R^{(2)}(\pi,0)$  \\
	
	\hline
	12 & $U^{(12)}(\frac{\tau}{N_p})$  \\

	\hline
	13 & $R^{(1)}(\pi,0) \otimes R^{(2)}(\pi,0)$  \\
	\hline
	14 & $U^{(12)}(\frac{\tau}{N_p})$  \\
	\hline
	15 & $R^{(1)}(\pi,\frac{4\pi}{5}) \otimes R^{(2)}(\pi,\frac{4\pi}{5})$  \\
	\hline
	16 & $U^{(12)}(\frac{\tau}{N_p})$  \\
	\hline
	17 & $R^{(1)}(\pi,\frac{2\pi}{5}) \otimes R^{(2)}(\pi,\frac{2\pi}{5})$  \\
	\hline
	18 & $U^{(12)}(\frac{\tau}{N_p})$  \\
	\hline
	19 & $R^{(1)}(\pi,\frac{4\pi}{5}) \otimes R^{(2)}(\pi,\frac{4\pi}{5})$  \\
	\hline
	20 & $U^{(12)}(\frac{\tau}{N_p})$  \\
	\hline
	21 & $R^{(1)}(\pi,0) \otimes R^{(2)}(\pi,0)$  \\
	\hline
	22 & $U^{(12)}(\frac{\tau}{2N_p})$  \\

	\hline \hline
	23 & $\mathds{I}$ or  $R^{(1)}(\frac{\pi}{2},\frac{\pi}{2})$ or $R^{(1)}(\frac{\pi}{2},0)$ \\
	\hline
	\end{tabular}
	\end{table}
	
\end{center}

\section{Reconstructed states}\label{sec:rec_states}
In the following, we determine the time evolution of the states $\rho_j=\lbrace \ket{0}\bra{0},\ket{1}\bra{1},\ket{+}\bra{+},\ket{\I}\bra{\I}\rbrace$ for different situations.

Reconstruction of the time evolution from the experimental data:
\begin{eqnarray}
\rho_1'&=&\left(\begin{array}{cc} 0.50 & 0.43+0.06\I \\ 0.43-0.06\I & 0.50 \end{array}\right) \label{eq:exprho1}\\
\rho_2'&=&\left(\begin{array}{cc} 0.41 & 0.42-0.11\I \\ 0.42+0.11\I & 0.59 \end{array}\right)\\
\rho_3'&=&\left(\begin{array}{cc} 0.51 & -0.05+0.06\I \\ -0.05-0.06\I & 0.49 \end{array}\right)\\
\rho_4'&=&\left(\begin{array}{cc} 0.04 & -0.03-0.07\I \\ -0.03+0.07\I & 0.96 \end{array}\right)\label{eq:exprho4}
\end{eqnarray}
and the corresponding eigenvalues $(0.94,0.06),(0.94,0.06),(0.41,0.58),(0.97,0.03)$.

To get a similar negativity of $\rho_\map$ solely by detuning without initial correlations (as an example we set $\rho_E=\ket{0}\bra{0}$) we assume a detuning of $\delta=0.4\Omega$ leading to $\lambda_-=-0.85$.The detuning influences the preparation, time evolution  and the measurement directions leading to:
\begin{eqnarray}
\rho_1'&=&\left(\begin{array}{cc}0.52  & 0.00 - 0.50\I \\ 0.00+ 0.50\I&  0.48\end{array}\right) \\
\rho_2'&=&\left(\begin{array}{cc}   0.72&  -0.26 + 0.35\I \\ -0.26 - 0.35\I &    0.28  \end{array}\right)\\
\rho_3'&=&\left(\begin{array}{cc} 0.14 & 0.34-0.02\I \\ 0.34+0.02\I & 0.86 \end{array}\right)\\
\rho_4'&=&\left(\begin{array}{cc} 0.15 & -0.36-0.02\I \\ -0.36+0.02\I & 0.85 \end{array}\right)
\end{eqnarray}
and the corresponding eigenvalues $(1.00,0.00),(0.99,0.01),(0.99,0.01),(1.01,-0.01)$. As can be seen, the behavior for solely detuning is quite different from that resulting from initial correlation. The main difference is that now all states stay nearly pure during the time evolution and that the reconstructed states themselves may have negative eigenvalues. Furthermore, the negativity is not detected by the witness used for initial correlations.

The difference between the theoretically predicted Choi-matrix and the experimentally reconstructed Choi-matrix can be the result of different errors such as (i) asymmetric detection error for the bright and the dark state, (ii) stray light shelving population from the $\ket{S_{1/2},F=1,m_f=+1}$ to the states $\ket{S_{1/2},F=1,m_f=0}$ or $\ket{S_{1/2},F=1,m_f=-1}$, or (iii) small detuning.

Detection errors shift the extremal expectation values $\langle \hat{z} \rangle =  \pm 1$ towards the average $\langle \hat{z} \rangle = 0$ (similar for $\langle \hat{z} \rangle$ and $\langle \hat{z} \rangle$). Asymmetric errors also shift the  zero-point of the expectation value towards the direction of smaller error. Typical errors in our experiments are $\varepsilon_B=0.06$ for the bright state (corresponding to $\ket{1},\ket{+},\ket{i}$) and $\varepsilon_D=0.03$ for the dark state. The matrix entries are directly proportional to the expectation values $\rho_{11}\sim \langle \hat{z} \rangle$, $\Re(\rho_{0,1})\sim \langle\sigma_x\rangle$ and $\Im(\rho_{0,1})\sim -\langle\hat{y}\rangle$. This leads to the theoretically predicted reconstructed states
\begin{eqnarray}
\rho_1'&=&\left(\begin{array}{cc}0.52  & 0.46 + 0.02\I \\ 0.46- 0.02\I&  0.48\end{array}\right) \\
\rho_2'&=&\left(\begin{array}{cc} 0.52  & 0.46 + 0.02\I \\ 0.46- 0.02\I&  0.48 \end{array}\right)\\
\rho_3'&=&\left(\begin{array}{cc} 0.52 & -0.02+0.02\I \\ -0.02-0.02\I & 0.48 \end{array}\right)\\
\rho_4'&=&\left(\begin{array}{cc} 0.06 & -0.02+0.02\I \\ -0.02-0.02\I & 0.94 \end{array}\right)
\end{eqnarray}

Stray light would lead to increased expectation values. The increase depends on the population in state $\ket{1}$ averaged over time. Here, we consider mainly the free evolution time, because the time for single qubit rotations (order of 10 $\mu$s) is very small in comparison. The states $\rho_j$ with $1\leq j \leq 3$ are always in the $xy$-plane during the conditional evolution time and therefore $\bar{p}(\ket{1})=0.5$. The state $\rho_4$ spends, due to the dynamical decoupling pulses, half of the time in $\ket{0}$ and half of the time in $\ket{1}$. Therefore, we find for this state also $\bar{p}(\ket{1})=0.5$. As a consequence, the effect of stray light is the same for all 4 input states. An assumed  population transfer of $5\%$ would lead to the following time evolutions:
\begin{eqnarray}
\rho_1'&=&\left(\begin{array}{cc}0.475  & 0.5 - 0.025\I \\ 0.5+ 0.025\I&  0.525\end{array}\right) \\
\rho_2'&=&\left(\begin{array}{cc} 0.475  & 0.5 - 0.025\I \\ 0.5+ 0.025\I&  0.525 \end{array}\right)\\
\rho_3'&=&\left(\begin{array}{cc} 0.475  & 0.025 - 0.025\I \\ 0.5+ 0.025\I&  0.525 \end{array}\right)\\
\rho_4'&=&\left(\begin{array}{cc} 0.00 & 0.025-0.025\I \\ 0.025+0.025\I & 1.00 \end{array}\right)
\end{eqnarray}

A detuning of $\delta=0.1 \Omega$ (influencing the preparation, time evolution, the dynamical decoupling and the measurement directions) together with the initial correlation of the system and its environment would lead to:
\begin{eqnarray}
\rho_1'&=&\left(\begin{array}{cc} 0.37 & 0.47-0.02\I \\ 0.47+0.02\I & 0.63 \end{array}\right) \\
\rho_2'&=&\left(\begin{array}{cc} 0.44 & 0.48+0.12\I \\ 0.48-0.12\I & 0.56 \end{array}\right)\\
\rho_3'&=&\left(\begin{array}{cc} 0.48 & 0.02+0.09\I \\ 0.02-0.09\I & 0.52 \end{array}\right)\\
\rho_4'&=&\left(\begin{array}{cc} 0.01 & -0.09-0.05\I \\ -0.09+0.05\I & 0.99 \end{array}\right) 
\end{eqnarray}
 A comparison with the experimentally reconstructed states \eq{eq:exprho1}-(\ref{eq:exprho4}) show remarkable similarities with different assumed errors for different states and measurements. However, this is not surprising since the experimental parameter such as the detuning or the intensity of the laser light used for state selective detection may fluctuate during a sequence of measurements. Therefore, not all errors are always present.

\end{appendix}

\bibliographystyle{apsrev}

\begin{thebibliography}{47}
\expandafter\ifx\csname natexlab\endcsname\relax\def\natexlab#1{#1}\fi
\expandafter\ifx\csname bibnamefont\endcsname\relax
  \def\bibnamefont#1{#1}\fi
\expandafter\ifx\csname bibfnamefont\endcsname\relax
  \def\bibfnamefont#1{#1}\fi
\expandafter\ifx\csname citenamefont\endcsname\relax
  \def\citenamefont#1{#1}\fi
\expandafter\ifx\csname url\endcsname\relax
  \def\url#1{\texttt{#1}}\fi
\expandafter\ifx\csname urlprefix\endcsname\relax\def\urlprefix{URL }\fi
\providecommand{\bibinfo}[2]{#2}
\providecommand{\eprint}[2][]{\url{#2}}

\bibitem[{\citenamefont{Weinstein et~al.}(2004)\citenamefont{Weinstein, Havel,
  Emerson, Boulant, Saraceno, Lloyd, and Cory}}]{Weinstein2004}
\bibinfo{author}{\bibfnamefont{Y.~S.} \bibnamefont{Weinstein}},
  \bibinfo{author}{\bibfnamefont{T.~F.} \bibnamefont{Havel}},
  \bibinfo{author}{\bibfnamefont{J.}~\bibnamefont{Emerson}},
  \bibinfo{author}{\bibfnamefont{N.}~\bibnamefont{Boulant}},
  \bibinfo{author}{\bibfnamefont{M.}~\bibnamefont{Saraceno}},
  \bibinfo{author}{\bibfnamefont{S.}~\bibnamefont{Lloyd}}, \bibnamefont{and}
  \bibinfo{author}{\bibfnamefont{D.~G.} \bibnamefont{Cory}},
  \bibinfo{journal}{J. Chem. Phys.} \textbf{\bibinfo{volume}{121}},
  \bibinfo{pages}{6117} (\bibinfo{year}{2004}).

\bibitem[{\citenamefont{Ringbauer et~al.}(2015)\citenamefont{Ringbauer, Wood,
  Modi, Gilchrist, White, and Fedrizzi}}]{Ringbauer2015}
\bibinfo{author}{\bibfnamefont{M.}~\bibnamefont{Ringbauer}},
  \bibinfo{author}{\bibfnamefont{C.~J.} \bibnamefont{Wood}},
  \bibinfo{author}{\bibfnamefont{K.}~\bibnamefont{Modi}},
  \bibinfo{author}{\bibfnamefont{A.}~\bibnamefont{Gilchrist}},
  \bibinfo{author}{\bibfnamefont{A.~G.} \bibnamefont{White}}, \bibnamefont{and}
  \bibinfo{author}{\bibfnamefont{A.}~\bibnamefont{Fedrizzi}},
  \bibinfo{journal}{Phys. Rev. Lett.} \textbf{\bibinfo{volume}{114}},
  \bibinfo{pages}{090402} (\bibinfo{year}{2015}).

\bibitem[{\citenamefont{Kuah et~al.}(2007)\citenamefont{Kuah, Modi,
  Rodriguez-Rosario, and Sudarshan}}]{Kuah2007}
\bibinfo{author}{\bibfnamefont{A.-M.} \bibnamefont{Kuah}},
  \bibinfo{author}{\bibfnamefont{K.}~\bibnamefont{Modi}},
  \bibinfo{author}{\bibfnamefont{C.~A.} \bibnamefont{Rodriguez-Rosario}},
  \bibnamefont{and} \bibinfo{author}{\bibfnamefont{E.~C.~G.}
  \bibnamefont{Sudarshan}}, \bibinfo{journal}{Phys. Rev. A}
  \textbf{\bibinfo{volume}{76}}, \bibinfo{pages}{042113}
  (\bibinfo{year}{2007}).

\bibitem[{\citenamefont{Pechukas}(1994)}]{Pechukas1994}
\bibinfo{author}{\bibfnamefont{P.}~\bibnamefont{Pechukas}},
  \bibinfo{journal}{Phys. Rev. Lett.} \textbf{\bibinfo{volume}{73}},
  \bibinfo{pages}{1060} (\bibinfo{year}{1994}).

\bibitem[{\citenamefont{Wood}(2009)}]{Wood2009}
\bibinfo{author}{\bibfnamefont{C.}~\bibnamefont{Wood}}, Ph.D. thesis,
  \bibinfo{school}{Macquarie University, Sydney, Australia}
  (\bibinfo{year}{2009}), \bibinfo{note}{arXiv:0911.3199}.

\bibitem[{\citenamefont{Modi and Sudarshan}(2010)}]{Modi2010}
\bibinfo{author}{\bibfnamefont{K.}~\bibnamefont{Modi}} \bibnamefont{and}
  \bibinfo{author}{\bibfnamefont{E.~C.~G.} \bibnamefont{Sudarshan}},
  \bibinfo{journal}{Phys. Rev. A} \textbf{\bibinfo{volume}{81}},
  \bibinfo{pages}{052119} (\bibinfo{year}{2010}).

\bibitem[{\citenamefont{Ziman}(2006)}]{Ziman2006}
\bibinfo{author}{\bibfnamefont{M.}~\bibnamefont{Ziman}} (\bibinfo{year}{2006}),
  \bibinfo{note}{arXiv:quant-ph/0603166}.

\bibitem[{\citenamefont{Knips et~al.}(2015)\citenamefont{Knips, Schwemmer,
  Klein, Reuter, T\'oth, and Weinfurter}}]{Knips2015}
\bibinfo{author}{\bibfnamefont{L.}~\bibnamefont{Knips}},
  \bibinfo{author}{\bibfnamefont{C.}~\bibnamefont{Schwemmer}},
  \bibinfo{author}{\bibfnamefont{N.}~\bibnamefont{Klein}},
  \bibinfo{author}{\bibfnamefont{J.}~\bibnamefont{Reuter}},
  \bibinfo{author}{\bibfnamefont{G.}~\bibnamefont{T\'oth}}, \bibnamefont{and}
  \bibinfo{author}{\bibfnamefont{H.}~\bibnamefont{Weinfurter}}
  (\bibinfo{year}{2015}), \bibinfo{note}{arXiv:1512.06866}.

\bibitem[{\citenamefont{Carteret and Terno}(2008)}]{Carteret2008}
\bibinfo{author}{\bibfnamefont{H.~A.} \bibnamefont{Carteret}} \bibnamefont{and}
  \bibinfo{author}{\bibfnamefont{K.}~\bibnamefont{Terno},
  \bibfnamefont{D.~R.and~Zyczkowski}}, \bibinfo{journal}{Phys. Rev. A}
  \textbf{\bibinfo{volume}{77}}, \bibinfo{pages}{042113}
  (\bibinfo{year}{2008}).

\bibitem[{\citenamefont{Vacchini and Amato}(2016)}]{Vacchini2016}
\bibinfo{author}{\bibfnamefont{B.}~\bibnamefont{Vacchini}} \bibnamefont{and}
  \bibinfo{author}{\bibfnamefont{G.}~\bibnamefont{Amato}},
  \bibinfo{journal}{Sci. Rep.} \textbf{\bibinfo{volume}{6}},
  \bibinfo{pages}{37328} (\bibinfo{year}{2016}).

\bibitem[{\citenamefont{Modi}(2012)}]{Modi2012}
\bibinfo{author}{\bibfnamefont{K.}~\bibnamefont{Modi}}, \bibinfo{journal}{Sci.
  Rep.} \textbf{\bibinfo{volume}{2}}, \bibinfo{pages}{581}
  (\bibinfo{year}{2012}).

\bibitem[{\citenamefont{Schwemmer et~al.}(2015)\citenamefont{Schwemmer, Knips,
  Richart, Moroder, Kleinmann, G\"uhne, and Weinfurter}}]{Schwemmer2015}
\bibinfo{author}{\bibfnamefont{C.}~\bibnamefont{Schwemmer}},
  \bibinfo{author}{\bibfnamefont{L.}~\bibnamefont{Knips}},
  \bibinfo{author}{\bibfnamefont{D.}~\bibnamefont{Richart}},
  \bibinfo{author}{\bibfnamefont{T.}~\bibnamefont{Moroder}},
  \bibinfo{author}{\bibfnamefont{M.}~\bibnamefont{Kleinmann}},
  \bibinfo{author}{\bibfnamefont{O.}~\bibnamefont{G\"uhne}}, \bibnamefont{and}
  \bibinfo{author}{\bibfnamefont{H.}~\bibnamefont{Weinfurter}},
  \bibinfo{journal}{Phys. Rev. Lett.} \textbf{\bibinfo{volume}{114}},
  \bibinfo{pages}{080403} (\bibinfo{year}{2015}).

\bibitem[{\citenamefont{Nielsen and Chuang}(2000)}]{Nielsen2000}
\bibinfo{author}{\bibfnamefont{M.~A.} \bibnamefont{Nielsen}} \bibnamefont{and}
  \bibinfo{author}{\bibfnamefont{I.~L.} \bibnamefont{Chuang}},
  \emph{\bibinfo{title}{Quantum Computation and Quantum Information}}
  (\bibinfo{publisher}{Cambridge University Press, Cambridge},
  \bibinfo{year}{2000}).

\bibitem[{\citenamefont{O'Brien et~al.}(2004)\citenamefont{O'Brien, Pryde,
  Gilchrist, James, Langford, Ralph, and White}}]{oBrien2004}
\bibinfo{author}{\bibfnamefont{J.~L.} \bibnamefont{O'Brien}},
  \bibinfo{author}{\bibfnamefont{G.~J.} \bibnamefont{Pryde}},
  \bibinfo{author}{\bibfnamefont{A.}~\bibnamefont{Gilchrist}},
  \bibinfo{author}{\bibfnamefont{D.~F.~V.} \bibnamefont{James}},
  \bibinfo{author}{\bibfnamefont{N.~K.} \bibnamefont{Langford}},
  \bibinfo{author}{\bibfnamefont{T.~C.} \bibnamefont{Ralph}}, \bibnamefont{and}
  \bibinfo{author}{\bibfnamefont{A.~G.} \bibnamefont{White}},
  \bibinfo{journal}{Phys. Rev. Lett.} \textbf{\bibinfo{volume}{93}},
  \bibinfo{pages}{080502} (\bibinfo{year}{2004}).

\bibitem[{\citenamefont{Mitchell et~al.}(2003)\citenamefont{Mitchell, Ellenor,
  Schneider, and Steinberg}}]{Mitchell2003}
\bibinfo{author}{\bibfnamefont{M.~W.} \bibnamefont{Mitchell}},
  \bibinfo{author}{\bibfnamefont{C.~W.} \bibnamefont{Ellenor}},
  \bibinfo{author}{\bibfnamefont{S.}~\bibnamefont{Schneider}},
  \bibnamefont{and} \bibinfo{author}{\bibfnamefont{A.~M.}
  \bibnamefont{Steinberg}}, \bibinfo{journal}{Phys. Rev. Lett.}
  \textbf{\bibinfo{volume}{91}}, \bibinfo{pages}{120402}
  (\bibinfo{year}{2003}).

\bibitem[{\citenamefont{Gessner et~al.}(2014)\citenamefont{Gessner, Ramm,
  Pruttivarasin, Buchleitner, Breuer, and H\"{a}ffner}}]{Gessner2014}
\bibinfo{author}{\bibfnamefont{M.}~\bibnamefont{Gessner}},
  \bibinfo{author}{\bibfnamefont{M.}~\bibnamefont{Ramm}},
  \bibinfo{author}{\bibfnamefont{T.}~\bibnamefont{Pruttivarasin}},
  \bibinfo{author}{\bibfnamefont{A.}~\bibnamefont{Buchleitner}},
  \bibinfo{author}{\bibfnamefont{H.-P.} \bibnamefont{Breuer}},
  \bibnamefont{and}
  \bibinfo{author}{\bibfnamefont{H.}~\bibnamefont{H\"{a}ffner}},
  \bibinfo{journal}{Nat. Phys.} \textbf{\bibinfo{volume}{10}},
  \bibinfo{pages}{105} (\bibinfo{year}{2014}).

\bibitem[{\citenamefont{Bagan et~al.}(2003)\citenamefont{Bagan, Baig, and
  Mu\~noz Tapia}}]{Bagan2003}
\bibinfo{author}{\bibfnamefont{E.}~\bibnamefont{Bagan}},
  \bibinfo{author}{\bibfnamefont{M.}~\bibnamefont{Baig}}, \bibnamefont{and}
  \bibinfo{author}{\bibfnamefont{R.}~\bibnamefont{Mu\~noz Tapia}},
  \bibinfo{journal}{Phys. Rev. A} \textbf{\bibinfo{volume}{67}},
  \bibinfo{pages}{014303} (\bibinfo{year}{2003}).

\bibitem[{\citenamefont{Gross et~al.}(2010)\citenamefont{Gross, Liu, Flammia,
  Becker, and Eisert}}]{Gross2010}
\bibinfo{author}{\bibfnamefont{D.}~\bibnamefont{Gross}},
  \bibinfo{author}{\bibfnamefont{Y.-K.} \bibnamefont{Liu}},
  \bibinfo{author}{\bibfnamefont{S.~T.} \bibnamefont{Flammia}},
  \bibinfo{author}{\bibfnamefont{S.}~\bibnamefont{Becker}}, \bibnamefont{and}
  \bibinfo{author}{\bibfnamefont{J.}~\bibnamefont{Eisert}},
  \bibinfo{journal}{Phys. Rev. Lett.} \textbf{\bibinfo{volume}{105}},
  \bibinfo{pages}{150401} (\bibinfo{year}{2010}).

\bibitem[{\citenamefont{Shabani et~al.}(2011)\citenamefont{Shabani, Kosut,
  Mohseni, Rabitz, Broome, Almeida, Fedrizzi, and White}}]{Shabani2011}
\bibinfo{author}{\bibfnamefont{A.}~\bibnamefont{Shabani}},
  \bibinfo{author}{\bibfnamefont{R.~L.} \bibnamefont{Kosut}},
  \bibinfo{author}{\bibfnamefont{M.}~\bibnamefont{Mohseni}},
  \bibinfo{author}{\bibfnamefont{H.}~\bibnamefont{Rabitz}},
  \bibinfo{author}{\bibfnamefont{M.~A.} \bibnamefont{Broome}},
  \bibinfo{author}{\bibfnamefont{M.~P.} \bibnamefont{Almeida}},
  \bibinfo{author}{\bibfnamefont{A.}~\bibnamefont{Fedrizzi}}, \bibnamefont{and}
  \bibinfo{author}{\bibfnamefont{A.~G.} \bibnamefont{White}},
  \bibinfo{journal}{Phys. Rev. Lett.} \textbf{\bibinfo{volume}{106}},
  \bibinfo{pages}{100401} (\bibinfo{year}{2011}).

\bibitem[{\citenamefont{Flammia and Liu}(2011)}]{Flammia2011}
\bibinfo{author}{\bibfnamefont{S.~T.} \bibnamefont{Flammia}} \bibnamefont{and}
  \bibinfo{author}{\bibfnamefont{Y.-K.} \bibnamefont{Liu}},
  \bibinfo{journal}{Phys. Rev. Lett.} \textbf{\bibinfo{volume}{106}},
  \bibinfo{pages}{230501} (\bibinfo{year}{2011}).

\bibitem[{\citenamefont{da~Silva et~al.}(2011)\citenamefont{da~Silva,
  Landon-Cardinal, and Poulin}}]{Silva2011}
\bibinfo{author}{\bibfnamefont{M.~P.} \bibnamefont{da~Silva}},
  \bibinfo{author}{\bibfnamefont{O.}~\bibnamefont{Landon-Cardinal}},
  \bibnamefont{and} \bibinfo{author}{\bibfnamefont{D.}~\bibnamefont{Poulin}},
  \bibinfo{journal}{Phys. Rev. Lett.} \textbf{\bibinfo{volume}{107}},
  \bibinfo{pages}{210404} (\bibinfo{year}{2011}).

\bibitem[{\citenamefont{Steffen et~al.}(2012)\citenamefont{Steffen, da~Silva,
  Fedorov, Baur, and Wallraff}}]{Steffen2012}
\bibinfo{author}{\bibfnamefont{L.}~\bibnamefont{Steffen}},
  \bibinfo{author}{\bibfnamefont{M.~P.} \bibnamefont{da~Silva}},
  \bibinfo{author}{\bibfnamefont{A.}~\bibnamefont{Fedorov}},
  \bibinfo{author}{\bibfnamefont{M.}~\bibnamefont{Baur}}, \bibnamefont{and}
  \bibinfo{author}{\bibfnamefont{A.}~\bibnamefont{Wallraff}},
  \bibinfo{journal}{Phys. Rev. Lett.} \textbf{\bibinfo{volume}{108}},
  \bibinfo{pages}{260506} (\bibinfo{year}{2012}).

\bibitem[{\citenamefont{Hofmann}(2005)}]{Hofmann2005}
\bibinfo{author}{\bibfnamefont{H.~F.} \bibnamefont{Hofmann}},
  \bibinfo{journal}{Phys. Rev. Lett.} \textbf{\bibinfo{volume}{94}},
  \bibinfo{pages}{160504} (\bibinfo{year}{2005}).

\bibitem[{\citenamefont{Gao et~al.}(2010)\citenamefont{Gao, Xu, Yao, G\"uhne,
  Cabello, Peng, Chen, and Pan}}]{Gao2010}
\bibinfo{author}{\bibfnamefont{W.-B.} \bibnamefont{Gao}},
  \bibinfo{author}{\bibfnamefont{P.}~\bibnamefont{Xu}},
  \bibinfo{author}{\bibfnamefont{X.-C.} \bibnamefont{Yao}},
  \bibinfo{author}{\bibfnamefont{O.}~\bibnamefont{G\"uhne}},
  \bibinfo{author}{\bibfnamefont{C.-Y.} \bibnamefont{Cabello},
  \bibfnamefont{G.~A.~Lu}}, \bibinfo{author}{\bibfnamefont{C.-Z.}
  \bibnamefont{Peng}}, \bibinfo{author}{\bibfnamefont{Z.-B.}
  \bibnamefont{Chen}}, \bibnamefont{and} \bibinfo{author}{\bibfnamefont{J.-W.}
  \bibnamefont{Pan}}, \bibinfo{journal}{Phys. Rev. Lett.}
  \textbf{\bibinfo{volume}{104}}, \bibinfo{pages}{020501}
  (\bibinfo{year}{2010}).

\bibitem[{\citenamefont{Orieux et~al.}(2013)\citenamefont{Orieux, Sansoni,
  Persechino, Mataloni, Rossi, and Macchiavello}}]{Orieux2013}
\bibinfo{author}{\bibfnamefont{A.}~\bibnamefont{Orieux}},
  \bibinfo{author}{\bibfnamefont{L.}~\bibnamefont{Sansoni}},
  \bibinfo{author}{\bibfnamefont{M.}~\bibnamefont{Persechino}},
  \bibinfo{author}{\bibfnamefont{P.}~\bibnamefont{Mataloni}},
  \bibinfo{author}{\bibfnamefont{M.}~\bibnamefont{Rossi}}, \bibnamefont{and}
  \bibinfo{author}{\bibfnamefont{C.}~\bibnamefont{Macchiavello}},
  \bibinfo{journal}{Phys. Rev. Lett.} \textbf{\bibinfo{volume}{111}},
  \bibinfo{pages}{220501} (\bibinfo{year}{2013}).

\bibitem[{\citenamefont{Choi}(1975)}]{Choi1975}
\bibinfo{author}{\bibfnamefont{M.-D.} \bibnamefont{Choi}},
  \bibinfo{journal}{Linear Algebra Appl.} \textbf{\bibinfo{volume}{10}},
  \bibinfo{pages}{285} (\bibinfo{year}{1975}).

\bibitem[{\citenamefont{Jiang et~al.}(2013)\citenamefont{Jiang, Luo, and
  Fu}}]{Jiang2013}
\bibinfo{author}{\bibfnamefont{M.}~\bibnamefont{Jiang}},
  \bibinfo{author}{\bibfnamefont{S.}~\bibnamefont{Luo}}, \bibnamefont{and}
  \bibinfo{author}{\bibfnamefont{S.}~\bibnamefont{Fu}}, \bibinfo{journal}{Phys.
  Rev. A} \textbf{\bibinfo{volume}{87}}, \bibinfo{pages}{022310}
  (\bibinfo{year}{2013}).

\bibitem[{\citenamefont{Ziman}(2008)}]{Ziman2008}
\bibinfo{author}{\bibfnamefont{M.}~\bibnamefont{Ziman}},
  \bibinfo{journal}{Phys. Rev. A} \textbf{\bibinfo{volume}{77}},
  \bibinfo{pages}{062112} (\bibinfo{year}{2008}).

\bibitem[{\citenamefont{Schmid et~al.}(2018)\citenamefont{Schmid, Ried, and
  Spekkens}}]{Schmid2018}
\bibinfo{author}{\bibfnamefont{D.}~\bibnamefont{Schmid}},
  \bibinfo{author}{\bibfnamefont{K.}~\bibnamefont{Ried}}, \bibnamefont{and}
  \bibinfo{author}{\bibfnamefont{R.~W.} \bibnamefont{Spekkens}}
  (\bibinfo{year}{2018}), \bibinfo{note}{arXiv:1806.02381}.

\bibitem[{\citenamefont{Moroder et~al.}(2013)\citenamefont{Moroder, Kleinmann,
  Schindler, Monz, G\"uhne, and Blatt}}]{Moroder2013}
\bibinfo{author}{\bibfnamefont{T.}~\bibnamefont{Moroder}},
  \bibinfo{author}{\bibfnamefont{M.}~\bibnamefont{Kleinmann}},
  \bibinfo{author}{\bibfnamefont{P.}~\bibnamefont{Schindler}},
  \bibinfo{author}{\bibfnamefont{T.}~\bibnamefont{Monz}},
  \bibinfo{author}{\bibfnamefont{O.}~\bibnamefont{G\"uhne}}, \bibnamefont{and}
  \bibinfo{author}{\bibfnamefont{R.}~\bibnamefont{Blatt}},
  \bibinfo{journal}{Phys. Rev. Lett.} \textbf{\bibinfo{volume}{110}},
  \bibinfo{pages}{180401} (\bibinfo{year}{2013}).

\bibitem[{\citenamefont{Hoeffding}(1963{\natexlab{a}})}]{Hoeffding1963}
\bibinfo{author}{\bibfnamefont{W.}~\bibnamefont{Hoeffding}},
  \bibinfo{journal}{J. Am. Stat. Assoc.} \textbf{\bibinfo{volume}{58}},
  \bibinfo{pages}{301} (\bibinfo{year}{1963}{\natexlab{a}}).

\bibitem[{\citenamefont{Knight}(1999)}]{Knight}
\bibinfo{author}{\bibfnamefont{K.}~\bibnamefont{Knight}},
  \emph{\bibinfo{title}{Mathematical Statistics (Chapman \& Hall/CRC Texts in
  Statistical Science)}} (\bibinfo{publisher}{Chapman and Hall/CRC},
  \bibinfo{year}{1999}), ISBN \bibinfo{isbn}{158488178X}.

\bibitem[{\citenamefont{Hjortrup et~al.}(2016)\citenamefont{Hjortrup, Haase,
  Wetterslev, and Perner}}]{Hjortrup2016}
\bibinfo{author}{\bibfnamefont{P.}~\bibnamefont{Hjortrup}},
  \bibinfo{author}{\bibfnamefont{N.}~\bibnamefont{Haase}},
  \bibinfo{author}{\bibfnamefont{J.}~\bibnamefont{Wetterslev}},
  \bibnamefont{and} \bibinfo{author}{\bibfnamefont{A.}~\bibnamefont{Perner}},
  \bibinfo{journal}{Crit. Care Resusc.} \textbf{\bibinfo{volume}{18}},
  \bibinfo{pages}{55} (\bibinfo{year}{2016}).

\bibitem[{\citenamefont{Khromova et~al.}(2012)\citenamefont{Khromova, Piltz,
  Scharfenberger, Gloger, Johanning, Var\'on, and Wunderlich}}]{Khromova2012}
\bibinfo{author}{\bibfnamefont{A.}~\bibnamefont{Khromova}},
  \bibinfo{author}{\bibfnamefont{C.}~\bibnamefont{Piltz}},
  \bibinfo{author}{\bibfnamefont{B.}~\bibnamefont{Scharfenberger}},
  \bibinfo{author}{\bibfnamefont{T.~F.} \bibnamefont{Gloger}},
  \bibinfo{author}{\bibfnamefont{M.}~\bibnamefont{Johanning}},
  \bibinfo{author}{\bibfnamefont{A.~F.} \bibnamefont{Var\'on}},
  \bibnamefont{and}
  \bibinfo{author}{\bibfnamefont{C.}~\bibnamefont{Wunderlich}},
  \bibinfo{journal}{Phys. Rev. Lett.} \textbf{\bibinfo{volume}{108}},
  \bibinfo{pages}{220502} (\bibinfo{year}{2012}).

\bibitem[{\citenamefont{W\"olk et~al.}(2015)\citenamefont{W\"olk, Piltz,
  Sriarunothai, and Wunderlich}}]{Woelk2015}
\bibinfo{author}{\bibfnamefont{S.}~\bibnamefont{W\"olk}},
  \bibinfo{author}{\bibfnamefont{C.}~\bibnamefont{Piltz}},
  \bibinfo{author}{\bibfnamefont{T.}~\bibnamefont{Sriarunothai}},
  \bibnamefont{and}
  \bibinfo{author}{\bibfnamefont{C.}~\bibnamefont{Wunderlich}},
  \bibinfo{journal}{J. Phys. B} \textbf{\bibinfo{volume}{48}},
  \bibinfo{pages}{075101} (\bibinfo{year}{2015}).

\bibitem[{\citenamefont{Piltz et~al.}(2016)\citenamefont{Piltz, Sriarunothai,
  Ivanov, W\"olk, and Wunderlich}}]{Piltz2016}
\bibinfo{author}{\bibfnamefont{C.}~\bibnamefont{Piltz}},
  \bibinfo{author}{\bibfnamefont{T.}~\bibnamefont{Sriarunothai}},
  \bibinfo{author}{\bibfnamefont{S.}~\bibnamefont{Ivanov}},
  \bibinfo{author}{\bibfnamefont{S.}~\bibnamefont{W\"olk}}, \bibnamefont{and}
  \bibinfo{author}{\bibfnamefont{C.}~\bibnamefont{Wunderlich}},
  \bibinfo{journal}{Sci. Adv.} \textbf{\bibinfo{volume}{2}},
  \bibinfo{pages}{e1600093} (\bibinfo{year}{2016}).

\bibitem[{\citenamefont{Sriarunothai et~al.}(2017)\citenamefont{Sriarunothai,
  Giri, W\"olk, and Wunderlich}}]{Sriarunothai2018}
\bibinfo{author}{\bibfnamefont{T.}~\bibnamefont{Sriarunothai}},
  \bibinfo{author}{\bibfnamefont{G.~S.} \bibnamefont{Giri}},
  \bibinfo{author}{\bibfnamefont{S.}~\bibnamefont{W\"olk}}, \bibnamefont{and}
  \bibinfo{author}{\bibfnamefont{C.}~\bibnamefont{Wunderlich}},
  \bibinfo{journal}{J. Mod. Opt.} \textbf{\bibinfo{volume}{65}},
  \bibinfo{pages}{560} (\bibinfo{year}{2017}).

\bibitem[{\citenamefont{Wunderlich}(2002)}]{Wunderlich2002}
\bibinfo{author}{\bibfnamefont{C.}~\bibnamefont{Wunderlich}},
  \emph{\bibinfo{title}{Conditional Spin Resonance with Trapped Ions}}
  (\bibinfo{publisher}{Springer Berlin Heidelberg}, \bibinfo{address}{Berlin,
  Heidelberg}, \bibinfo{year}{2002}), pp. \bibinfo{pages}{261--273}, ISBN
  \bibinfo{isbn}{978-3-662-04897-9}.

\bibitem[{\citenamefont{W\"olk and Wunderlich}(2017)}]{Woelk2017}
\bibinfo{author}{\bibfnamefont{S.}~\bibnamefont{W\"olk}} \bibnamefont{and}
  \bibinfo{author}{\bibfnamefont{C.}~\bibnamefont{Wunderlich}},
  \bibinfo{journal}{New J. Phys.} \textbf{\bibinfo{volume}{19}},
  \bibinfo{pages}{083021} (\bibinfo{year}{2017}).

\bibitem[{\citenamefont{Piltz et~al.}(2013)\citenamefont{Piltz, Scharfenberger,
  Khromova, Var\'on, and Wunderlich}}]{Piltz2013}
\bibinfo{author}{\bibfnamefont{C.}~\bibnamefont{Piltz}},
  \bibinfo{author}{\bibfnamefont{B.}~\bibnamefont{Scharfenberger}},
  \bibinfo{author}{\bibfnamefont{A.}~\bibnamefont{Khromova}},
  \bibinfo{author}{\bibfnamefont{A.~F.} \bibnamefont{Var\'on}},
  \bibnamefont{and}
  \bibinfo{author}{\bibfnamefont{C.}~\bibnamefont{Wunderlich}},
  \bibinfo{journal}{Phys. Rev. Lett.} \textbf{\bibinfo{volume}{110}},
  \bibinfo{pages}{200501} (\bibinfo{year}{2013}).

\bibitem[{\citenamefont{Genov et~al.}(2017)\citenamefont{Genov, Daniel,
  Vitanov, and Halfmann}}]{Genov2016}
\bibinfo{author}{\bibfnamefont{G.~T.} \bibnamefont{Genov}},
  \bibinfo{author}{\bibfnamefont{S.}~\bibnamefont{Daniel}},
  \bibinfo{author}{\bibfnamefont{N.~V.} \bibnamefont{Vitanov}},
  \bibnamefont{and} \bibinfo{author}{\bibfnamefont{T.}~\bibnamefont{Halfmann}},
  \bibinfo{journal}{Phys. Rev. Lett.} \textbf{\bibinfo{volume}{118}},
  \bibinfo{pages}{133202} (\bibinfo{year}{2017}).

\bibitem[{\citenamefont{van Enk and Blume-Kohout}(2013)}]{Enk2013}
\bibinfo{author}{\bibfnamefont{S.~J.} \bibnamefont{van Enk}} \bibnamefont{and}
  \bibinfo{author}{\bibfnamefont{R.}~\bibnamefont{Blume-Kohout}},
  \bibinfo{journal}{New J. Phys.} \textbf{\bibinfo{volume}{15}},
  \bibinfo{pages}{025024} (\bibinfo{year}{2013}).

\bibitem[{\citenamefont{Gu{\c{t}}{\u{a}}
  et~al.}(2012)\citenamefont{Gu{\c{t}}{\u{a}}, Kypraios, and
  Dryden}}]{Guta2012}
\bibinfo{author}{\bibfnamefont{M.}~\bibnamefont{Gu{\c{t}}{\u{a}}}},
  \bibinfo{author}{\bibfnamefont{T.}~\bibnamefont{Kypraios}}, \bibnamefont{and}
  \bibinfo{author}{\bibfnamefont{I.}~\bibnamefont{Dryden}},
  \bibinfo{journal}{New J. Phys.} \textbf{\bibinfo{volume}{14}},
  \bibinfo{pages}{105002} (\bibinfo{year}{2012}).

\bibitem[{\citenamefont{Laine et~al.}(2010)\citenamefont{Laine, Piilo, and
  Breuer}}]{Laine2010}
\bibinfo{author}{\bibfnamefont{E.-M.} \bibnamefont{Laine}},
  \bibinfo{author}{\bibfnamefont{J.}~\bibnamefont{Piilo}}, \bibnamefont{and}
  \bibinfo{author}{\bibfnamefont{H.-P.} \bibnamefont{Breuer}},
  \bibinfo{journal}{Europhys. Lett.} \textbf{\bibinfo{volume}{92}},
  \bibinfo{pages}{60010} (\bibinfo{year}{2010}).

\bibitem[{\citenamefont{Gessner and Breuer}(2011)}]{Gessner2011}
\bibinfo{author}{\bibfnamefont{M.}~\bibnamefont{Gessner}} \bibnamefont{and}
  \bibinfo{author}{\bibfnamefont{H.-P.} \bibnamefont{Breuer}},
  \bibinfo{journal}{Phys. Rev. Lett.} \textbf{\bibinfo{volume}{107}},
  \bibinfo{pages}{180402} (\bibinfo{year}{2011}).

\bibitem[{\citenamefont{Rossatto et~al.}(2011)\citenamefont{Rossatto, Werlang,
  Castelano, Villas-Boas, and Fanchini}}]{Rossatto2011}
\bibinfo{author}{\bibfnamefont{D.~Z.} \bibnamefont{Rossatto}},
  \bibinfo{author}{\bibfnamefont{T.}~\bibnamefont{Werlang}},
  \bibinfo{author}{\bibfnamefont{L.~K.} \bibnamefont{Castelano}},
  \bibinfo{author}{\bibfnamefont{C.~J.} \bibnamefont{Villas-Boas}},
  \bibnamefont{and} \bibinfo{author}{\bibfnamefont{F.~F.}
  \bibnamefont{Fanchini}}, \bibinfo{journal}{Phys. Rev. A}
  \textbf{\bibinfo{volume}{84}}, \bibinfo{pages}{042113}
  (\bibinfo{year}{2011}).

\bibitem[{\citenamefont{Hoeffding}(1963{\natexlab{b}})}]{Hoeffding}
\bibinfo{author}{\bibfnamefont{W.}~\bibnamefont{Hoeffding}},
  \bibinfo{journal}{J. Am. Stat. Assoc.} \textbf{\bibinfo{volume}{58}},
  \bibinfo{pages}{13} (\bibinfo{year}{1963}{\natexlab{b}}).

\end{thebibliography}

\end{document}